\documentclass[aps,pre,reprint,superscriptaddress]{revtex4-1}
\usepackage[utf8]{inputenc}
\usepackage{graphicx}
\usepackage{placeins}
\usepackage{floatflt}
\usepackage{natbib}
\usepackage{epsfig,epsf,rotating}
\usepackage{amsmath}
\graphicspath{ {figures/} }

\begin{document}
\bibliographystyle{apsrev4-1}
\title{Demonstrating Universal Scaling in Quench Dynamics of a Yukawa One-Component Plasma}
\author{T.K. Langin}
\email[]{tkl1@rice.edu}
\affiliation{Rice University, Department of Physics and Astronomy, Houston, Texas USA}
\author{T. Strickler}
\affiliation{Rice University, Department of Physics and Astronomy, Houston, Texas USA}
\author{N. Maksimovic}
\affiliation{Rice University, Department of Physics and Astronomy, Houston, Texas USA}
\author{P. McQuillen}
\affiliation{Rice University, Department of Physics and Astronomy, Houston, Texas USA}
\author{T. Pohl}
\affiliation{Max Planck Institute for Complex Systems, Dresden, Germany}
\author{D. Vrinceanu}
\affiliation{Texas Southern University, Department of Physics, Houston, Texas, USA}
\author{T.C. Killian}
\affiliation{Rice University, Department of Physics and Astronomy, Houston, Texas USA}

\date{\today}

\begin{abstract}
 The Yukawa one-component plasma (OCP) is a paradigm model for describing plasmas that contain one component of interest and one or more other components that can be treated as a neutralizing, screening background.  In appropriately scaled units, interactions are characterized entirely by a screening parameter, $\kappa$.  As a result, systems of similar $\kappa$ show the same dynamics, regardless of the underlying parameters (e.g., density and temperature).  We demonstrate this behavior using ultracold neutral plasmas (UNP) created by photoionizing a cold ($T\le10$\,mK) gas.  The ions in UNP systems are well described by the Yukawa model, with the electrons providing the screening.  Creation of the plasma through photoionization can be thought of as a rapid quench from $\kappa_{0}=\infty$ to a final $\kappa$ value set by the electron density and temperature.  We demonstrate experimentally that the post-quench dynamics are universal in $\kappa$ over a factor of 30 in density and an order of magnitude in 
temperature.  Results are compared with molecular dynamics simulations.  We also demonstrate that features of the post-quench kinetic energy evolution, such as disorder-induced heating and kinetic-energy oscillations, can be used to determine the plasma density and the electron temperature.
\end{abstract}


\maketitle

\section{Introduction}
\label{Intro}

The Yukawa one-component-plasma (OCP) model, in which particles interact through a screened, repulsive $1/r$ potential (Eq.~\ref{YukawaInteraction}), is used to describe systems such as the cores of white dwarf stars \cite{salpeter:WhiteDwarf} and Jovian planets \cite{stevenson:Jovian,remington:Astro}, plasmas produced during inertial confinement fusion \cite{lindl:ICF}, dusty plasmas consisting of highly charged dust particles \cite{Melzer:CorrReCryst,morfill:Dusty}, charge-stabilized colloidal systems such as latex spheres in a polar solvent \cite{wojtowicz:ColloidalMelting,grest:YukawaModelColloids}, and ions in ultracold neutral plasmas (UNPs) \cite{killian:UNPCreate,rolston:UNPCreate}, which are the focus of this work.  Such systems contain one species of interest (e.g. ions) and at least one other species (e.g. electrons, polar molecules) that acts to screen interactions between particles of the species of interest.  In addition to describing many real systems, this model is also used in molecular 
dynamics studies of strongly coupled plasmas \cite{Hamaguchi:UCalc,hamaguchi:YukawaMD,murillo:UFast}, and for research on phase transitions \cite{grest:YukawaModelColloids}.  Interactions in the Yukawa model take the form:

\begin{equation}
  V_{ij}(r_{i},r_{j})=\frac{U_{0}}{r_{ij}}\exp\left(-\frac{r_{ij}}{\lambda_{D}}\right)
 \label{YukawaInteraction}
\end{equation}

\noindent where $r_{ij}=|\vec{r_{i}}-\vec{r_{j}}|$, $U_{0}$ is a measure of the interaction strength, and $\lambda_{D}$ is known as the Debye screening length.  

One important feature of the Yukawa model is that, in suitably normalized units, dynamics depend solely on the screening parameter, $\kappa=\lambda_{D}/a$, where $a=\left[3/\left(4\pi n\right)\right]^{1/3}$ is the Wigner-Seitz radius and $n$ is the density.  This universal scaling follows from the classical scaling invariance for a system of charged particles, and it allows for comparisons to be made between systems with different $n$, $U_{0}$, and $T$, where $T$ is the temperature.  

In this paper, we demonstrate universal scaling in the ion component of UNPs created by photoionizing a laser cooled ($T\le10$\,mK) magneto-optically trapped gas \cite{killian:UNPCreate,rolston:UNPCreate}.  Ion interactions in UNPs are well described by the Yukawa model, with $U_{0}=e^{2}/(4\pi\epsilon_{0})$.  Electrons serve as a neutralizing and screening background, with Debye screening length $\lambda_{D}=\sqrt{k_{B}T_{e}\epsilon_{0}/(ne^{2})}$.  This leads to the normalized units chosen here of energy $E\rightarrow E/(e^{2}/(4\pi\epsilon_{0}a))$, position $r\rightarrow r/a$, and time $t\rightarrow \omega_{pi}t$, where the time scaling factor is the ion plasma oscillation frequency $\omega_{pi}=\sqrt{ne^{2}/(\epsilon_{0}m_{i})}$.

Universal scaling is demonstrated by measuring the evolution of the ion kinetic energy after photoionization, which can be thought of as a rapid quench from $\kappa_{0}=\infty$ (i.e. the non-interacting gas) to $\kappa_{F}$, where $\kappa_{F}$ is an experimentally adjustable parameter (see Sec.~\ref{UNPs}).  We verify that, in appropriately scaled units, the post-quench kinetic energy evolution in plasmas with the same value of $\kappa_{F}$ are identical even if the density is varied over a factor of 30 ($n\sim3\times10^{14}-9\times10^{15}$m$^{-3}$) and the electron temperature is varied over an order of magnitude ($T_{e}=50-450$\,K).  We compare all of our results to molecular dynamics (MD) simulations.  In addition to the specific application to plasma physics, this work exploits the Yukawa OCP as a paradigm model to explore the dynamics of many-body systems far from equilibrium, which is of interest in many areas of science.  The scaling of the dynamics of three-body recombination in an ultracold 
neutral plasma was investigated with molecular dynamics simulations in~\cite{grant:TBRUnivScaling}.

One interesting feature of the quench of an infinitely screened Yukawa OCP is that it automatically results in a strongly coupled plasma (SCP) \cite{killian:UNPCreate}.  Coupling is parameterized by the ratio of the nearest neighbor Coulomb interaction energy to the kinetic energy

\begin{equation}
 \Gamma_{s}=\frac{\left(Ze\right)^{2}/\left(4\pi\epsilon_{0}a\right)}{k_{B}T_{s}},
 \label{CCP}
\end{equation}

\noindent where the ratio $\Gamma_{s}$ is the Coulomb coupling parameter for species $s$ at temperature $T_{s}$.  For strong coupling, $\Gamma\gtrsim1$.  UNPs equilibrate with $\Gamma_{i}=2-4$.  The electrons remain weakly coupled with $\Gamma_{e}\le0.1$ \cite{killian:electronTemp}.  The cores of white dwarf stars ($\Gamma=10-200$), the cores of Jovian planets ($\Gamma=20-50$), and plasmas produced in laser-implosion experiments, such as those designed to produce inertial confinement fusion (ICF) can also be near or in the strongly coupled regime \cite{ichimaru:SCPReview}.  

We note useful relations between parameters describing the electrons: $\Gamma_{e}=\kappa^{2}/3$, and the number of electrons per Debye sphere is $N_{D}=\kappa^{-3}$.  Thus, for the plasmas described here $\kappa\le0.55$ and $N_{D}\ge 6$.  For the remainder of the paper, we will only refer to Coulomb coupling parameters for the ions, and we will drop the subscript on $\Gamma$.

The rest of the paper is structured as follows: In Sec.~\ref{UNPs} we discuss prior studies of UNPs.  In Sec.~\ref{sec:Methods} we provide details of our experiment and MD simulation.  In Sec.~\ref{Results} we discuss the results of the joint experimental and numerical study.

\section{Ultracold Neutral Plasmas}
\label{UNPs}

UNPs can be generated by laser photoionization of either laser-cooled, magneto-optically-trapped gases of atoms \cite{killian:UNPCreate,rolston:UNPCreate} or molecular beams \cite{grant:UNPCreate}, or by spontaneous avalanche ionization in a dense gas of highly excited Rydberg atoms \cite{rolston:SpontIonization,gallagher:SpontIonization,weidemuller:SpontIonization}.  Typical UNP densities range from $n=10^{14}$\,m$^{-3}$ to $10^{17}$\,m$^{-3}$.  The ion temperatures, $T_{i}$, in UNPs can be as low as $\sim100$\,mK \cite{killian:UNPEvolve}.  The electron temperature, $T_{e}$, in UNPs generated by the photoionization process is determined by the excess photon energy above the photoionization threshold \cite{rolston:UNPCreate}.  The photon energy is a controllable parameter, granting control over $T_{e}$ and, consequently, $\kappa_{F}$.

The UNP dynamics after photoionization can be described by a sequence of events that take place on substantially different timescales.  First, on a timescale $\omega_{pe}^{-1}=\sqrt{\epsilon_{0}m_{e}/(ne^{2})}\sim10$\,ps$-1$\,ns, where $\omega_{pe}$ is the electron plasma frequency, the electrons equilibrate to close to a thermal distribution \cite{rolston:UNPCreate}.  If $\kappa\gtrsim0.55$, additional phenomena occur, such as three-body recombination (TBR) \cite{killian:electronTemp}, which can cause deviations from the simple description of UNP dynamics and from the Yukawa OCP model.  For this reason, we restrict ourselves to $\kappa\lesssim0.55$ in this work.

As the electrons equilibrate, the ions retain the kinetic energy distribution of the $\sim10$\,mK atoms.  If the ions were to equilibrate to this temperature, they would be deep into the SCP regime (for example, $\Gamma=580$ for typical values $n=10^{16}$\,m$^{-3}$ and $T_{i}=10$\,mK).  However, a heating process known as disorder-induced heating (DIH) reduces the achieved coupling to $\Gamma=2-4$ after equilibration.

\subsection{Disorder-Induced Heating and Kinetic-Energy Oscillations}
\label{DIH}

\begin{figure}[!h]
  \includegraphics{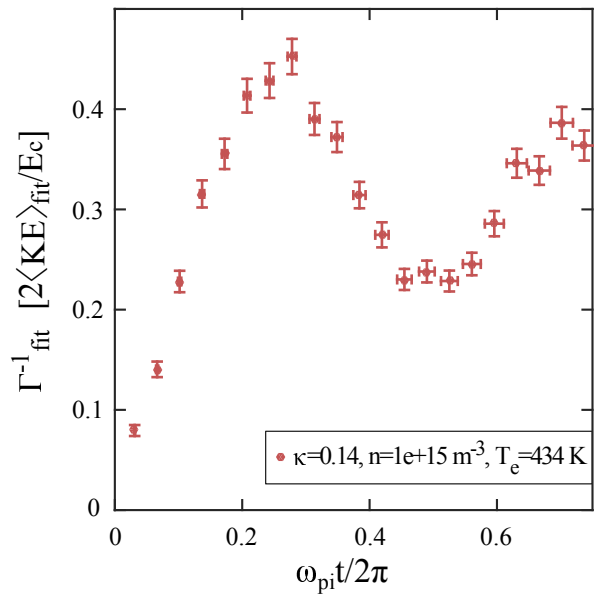}
\caption{Experimental DIH curve plotted in scaled units for average one-dimensional kinetic energy per ion $\langle KE\rangle_{fit}$ and time $t$.  Curve taken at $\kappa_{F}=0.14$, $n=10^{15}$\,m$^{-3}$, and $T_{e}=440$\,K, which yields $E_{c}/k_{B}=2.7$\,K and $2\pi/\omega_{pi}=1.4$\,$\mu$s}
\label{fig:DIHExample}
\end{figure}

After the electron equilibration, the ions undergo DIH on a timescale $\omega_{pi}^{-1}\sim100$\,ns$-1$\,$\mu$s, which defines the natural time scale for the ion dynamics \cite{killian:DIH,bergeson:DIH,bergeson:DIHScreen}.  During DIH, the ion kinetic energy first increases dramatically, then subsequently undergoes damped oscillations known as ``kinetic energy oscillations'' (KEOs), which occur at frequency $\sim2\omega_{pi}$.  Figure~\ref{fig:DIHExample} shows a DIH curve plotted in natural units, with time scaled by $2\pi/\omega_{pi}$ and approximate average one-dimensional kinetic energy per ion ($\langle KE\rangle_{fit}$) scaled by the nearest neighbor Coulomb energy, $E_{c}=e^{2}/(4\pi\epsilon_{0}a)$, which yields the inverse of an effective Coulomb coupling parameter, $\Gamma_{fit}^{-1}=2\langle KE\rangle_{fit}/E_{c}$.  We will describe in Sec.~\ref{sec:Methods} how this approximate measure of the kinetic energy is 
derived from the data.

DIH results from the fact that, before photoionization, the equilibrium positions of the atoms are uncorrelated.  After plasma creation, however, strong Coulomb interactions make close pairs of ions energetically unfavorable.  Spatial correlations develop as the ions move to reduce their potential energy.  By conservation of energy, this increases the average kinetic energy per ion, leading to an elevated temperature and reduced $\Gamma$ in equilibrium.  We model this process as equilibration of a Yukawa OCP after a quench of interactions from $\kappa=\infty$ to $\kappa_{F}$.  

DIH was predicted \cite{murillo:DIHPRL} soon after the first UNP experiments \cite{rolston:UNPCreate}. The magnitude of the resulting heating can be calculated by using the pair correlation function, $g(r)$, to determine the change in interaction energy.  The pair correlation function reflects how the local density near any particle (taken to be at the origin) is modified by correlations: $n_{local}(r)=g(r)n$ (e.g., in an uncorrelated system, $g(r)=1$).

At equilibrium, $g(r)$ is solely determined by $\Gamma$ and $\kappa$ \cite{Hamaguchi:UCalc,murillo:DIH,murillo:DIHPRL}.  The change in kinetic energy by the end of the equilibration is equivalent to the negative of the change in interaction energy, and therefore:

\begin{equation}
 \frac{k_{B}(T_{f}-T_{i})}{N}= -\frac{\Delta U_{c}}{N} = \frac{n}{2}\int V_{ij}(r)\left[1-g(r,\Gamma,\kappa)\right] d^{3}\vec{r}
 \label{corrEnergy}
\end{equation}

\noindent By taking advantage of the isotropy of the Yukawa interaction and expressing quantities in normalized units ($\tilde{r}=r/a$), and setting the initial temperature $T_{i}\approx0$, this takes the simpler form:

\begin{equation}
 \frac{k_{B}T_{f}}{e^{2}/(4\pi\epsilon_{0}a)}=\Gamma^{-1}(\kappa)=\int_{0}^{\infty}\tilde{r}\exp[-\kappa\tilde{r}]\left[1-g(\tilde{r},\Gamma,\kappa)\right]d\tilde{r}
 \label{DIHGammaSolve}
\end{equation}

\noindent where $\Gamma(\kappa)$ is the Coulomb coupling parameter after equilibration, and $\Gamma^{-1}$ can be viewed as a temperature or kinetic energy in scaled units.

\begin{figure}[!h]
  \includegraphics{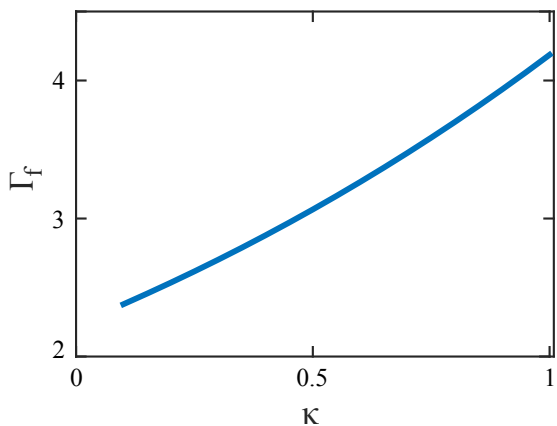}
\caption{$\Gamma(\kappa)$ from solving Eq.~\ref{DIHGammaSolve}.}
\label{fig:GamVKap}
\end{figure}

Already, we see evidence of universal scaling in the equilibration process, since Eq.~\ref{DIHGammaSolve} can be solved for $\Gamma$ as a function of $\kappa$ using MD simulation results for $g(\tilde{r},\Gamma,\kappa)$ (see Fig.~\ref{fig:GamVKap}) \cite{Hamaguchi:UCalc}.  Previous experiments have confirmed that the $\Gamma(\kappa)$ achieved after equilibration matches Eq.~\ref{DIHGammaSolve} \cite{killian:DIH,bergeson:DIHScreen}.

The KEOs, on the other hand, are not described by an analytical expression \cite{killian:DIH,murillo:UFast}.  MD simulations have shown that the frequency of these oscillations is approximately $2\omega_{pi}$.  Increasing $\kappa$, however, softens the ion-ion interaction and slows the oscillations slightly for the experimentally accessible range of screening \cite{bergeson:DIH}.  The KEOs result from the exchange between kinetic energy and interaction energy as $g(r)$ relaxes towards equilibrium \cite{murillo:UFast}. Oscillations at frequencies near $\omega_{pi}$ are a common feature in dynamics of SCPs; for example, they are observed in velocity auto-correlation functions \cite{Hamaguchi:VAFOsc,Killian:VAF,Hansen:NonEqRelax,Mazenko:NonEqRelax}.

However, the principle of universal scaling demands that the entire curve, not just the frequency and the post-equilibration coupling, depend solely on $\kappa_{F}$.  This dependence is established in Sec.~\ref{Results} over a factor of 30 in $n$ and an order of magnitude in $T_{e}$.

\section{Methods}
\label{sec:Methods}

\subsection{Experiment}
\label{Experiment}

We create UNPs by photoionizing a gas of laser-cooled $^{88}$Sr atoms in a magneto-optical trap (MOT) \cite{killian:UNPCreate}.  We use a two photon sequence to photoionize the gas: one photon of $461$\,nm from a pulsed-dye-amplified CW laser to excite the $^{1}$S$_{0}\rightarrow^{1}$P$_{1}$ transition and another tunable photon (405-413\,nm) from a pulsed dye laser to ionize from the $^{1}$P$_{1}$ state.  We refer to the latter as the ionization laser.  Both systems are pumped by 10 ns, 355 nm pulses from third harmonic generation of a pulsed Nd:YAG laser.

By tuning the wavelength of the ionization laser, we adjust the plasma electron temperature in the range $T_{e}=49-440$\,K \cite{killian:electronTemp}.  The plasma has a gaussian density profile, $n(r)=n_{0}\exp\left[-r^{2}/(2\sigma_{0}^{2})\right]$ with width $\sigma_{0}=1-2$\,mm and peak density $n_{0}=3\times 10^{14}-9\times 10^{15}$\,m$^{-3}$.  

The kinetic energy and density of the ion component are probed using laser induced fluorescence (LIF) spectroscopy at $\lambda=422$\,nm, corresponding to the $^{2}$S$_{1/2}\rightarrow^{2}$P$_{1/2}$ Sr$^{+}$ transition \cite{Killian:LIF}.  Fluorescence is excited in a 1\,mm thick sheet passing through the center of the plasma.  This ensures that fluorescence is excited in a region with little density variation along the unresolvable imaging axis (for our smallest width, $\sigma_{0}=1$\,mm, density varies by $e^{-1/8}$ along the image axis).  We define our coordinate system such that the LIF laser lies in the x-y plane, propagating along the x-axis, at approximately z=0.

A fraction of the fluorescence emitted perpendicular to the sheet is then collected and imaged via a 1:1 optical relay onto an intensified charge coupled device (ICCD) with a 13\,$\mu$m pixel size.  This allows for regional analysis of small volumes of roughly constant density.  The ICCD can be gated with a 30\,ns resolution, which allows for time-resolved measurements of the DIH curve. The DIH timescale is $t_{DIH}=2\pi/\omega_{pi}\ge500$\,ns for typical UNP densities.

We scan $f$, the frequency detuning of the LIF laser from resonance, to obtain a fluorescence spectrum.  The spectrum is a convolution of a lorentzian of width given by the sum of the laser and natural linewidths ($\gamma_{l}$=6 MHz and $\gamma_{n}$=20 MHz respectively) with Doppler shifts resulting from the velocity distribution, $D(v,x,y)$, for plasma at position (x,y):

\begin{equation}
 S(x,y,f) = C(x,y)\int_{-\infty}^{\infty}dv' \frac{D(v',x,y)}{\gamma^{2}/4+(f-\frac{v'}{\lambda})^{2}}
 \label{specArbDistro}
\end{equation}

\noindent where $\gamma=\gamma_{l}+\sqrt{1+s_{0}}\gamma_{n}$ is the lorentzian width with power broadening taken into account, $s_{0}$ is the saturation parameter for the transition, and $C(x,y)$ is proportional to the local density at $(x,y)$ and the overall photon detection efficiency \cite{Killian:LIF}.

If we assume that the velocity distribution can be described with a thermalized gaussian with width $\sigma_{v}(x,y)=\sqrt{k_{B}T_{fit}(x,y)/m_{i}}$, with $T_{fit}(x,y)$ being the local temperature, the signal can be expressed as:

\begin{equation}
 S(x,y,f)=\frac{C(x,y)}{\sqrt{2\pi\sigma_{v}(x,y)}}\int_{-\infty}^{\infty} \frac{\exp\left[-\frac{(v'-v_{0}(x))^{2}}{2\sigma_{v}(x,y)^{2}}\right]dv'}{\gamma^{2}/4+(f-\frac{v'}{\lambda})^{2}}.
 \label{specAssumeGauss}
\end{equation}

\noindent As we explain in Sec.~\ref{MDSim}, the assumption of a thermalized plasma is not strictly valid during DIH; the velocity distribution can differ slightly from a gaussian, especially during the initial rise in kinetic energy.  However, fitting the velocity distribution to a gaussian and obtaining a value of $T_{fit}$ provides a well-defined method of characterizing data from experiments and simulations, and it serves as our primary analysis tool.  While $T_{fit}$ does not always have a 1:1 correspondence to the average kinetic energy, we will use it to define an approximate kinetic energy in scaled units as $\Gamma_{fit}^{-1}=k_{B}T_{fit}/E_{c}$.  In equilibrium, $\Gamma_{fit}=\Gamma$.

Another source of complication is that $T_{fit}(x,y)$ will vary with density throughout the plasma during the DIH process, as both the frequency of the KEOs and the overall energy scaling are both density dependent.  Thus, we restrict our analysis to a region of area $1$\,mm$\times1$\,mm (80$\times$80 pixels) centered on the center of the plasma as a compromise between the desire to maximize signal and the desire to keep the density relatively uniform over the analysis region.

It is important to note that the plasma is unconfined, and that after it is created the electron thermal pressure causes it to expand radially.  This will cause the average velocity along the laser axis within a pixel to depend on $x$,  which we take into account by allowing for a bulk velocity in each region, $v_{0}(x)$, in Eq.~\ref{specAssumeGauss}.  The effect of expansion is discussed in greater detail in previous papers \cite{Killian:LIF,killian:UNPEvolve}.  For the work discussed in this paper the effect is relatively minor, as $t_{DIH}$ is shorter than $\tau_{exp}=\sqrt{k_{B}T_{e}\sigma_{0}^{2}/m_{i}}$, the timescale for expansion.  For all experiments discussed in this work, $t_{DIH}/\tau_{exp}\le0.1$.

We divide the 80$\times$80 pixel analysis region into 20 regions of size $80\times4$, with the short axis along $x$ in order to minimize spread in expansion velocity along $x$ within each region.  The spectrum in each region is then fit to Eq.~\ref{specAssumeGauss}, with $C$, $v_{0}$, and $\sigma_{v}$ as free parameters.  We then take the average of the temperatures derived from all 20 fits to $\sigma_{v}$.  The density, $n$, can be determined from the value of the amplitude, $C$, which is calibrated with absorption imaging measurements as described in \cite{killian:UNPCreate}.  The effect on the DIH curve of averaging temperatures from regions of differing density to determine $T_{fit}$ is discussed in detail in Sec.~\ref{nAndTeFit} and in the Appendix.

\subsection{Molecular Dynamics Simulations}
\label{MDSim}

The MD simulations evolve a Yukawa OCP of $N=20000$ particles in a cubic volume with periodic boundary conditions using the minimum image convention~\cite{smit:MinimumImage} and a leap-frog integrator \cite{eastwood:SimulationTheory} of Hamilton's equations of motion, expressed in natural units, with a timestep of $0.0035\omega_{pi}t$.  Simulations were performed with $\kappa$ ranging from 0.12 to 0.55, with 50 runs conducted for each chosen $\kappa$.  The initial conditions for the particles are random positions and zero kinetic energy.

\begin{figure}[!h]
  \includegraphics{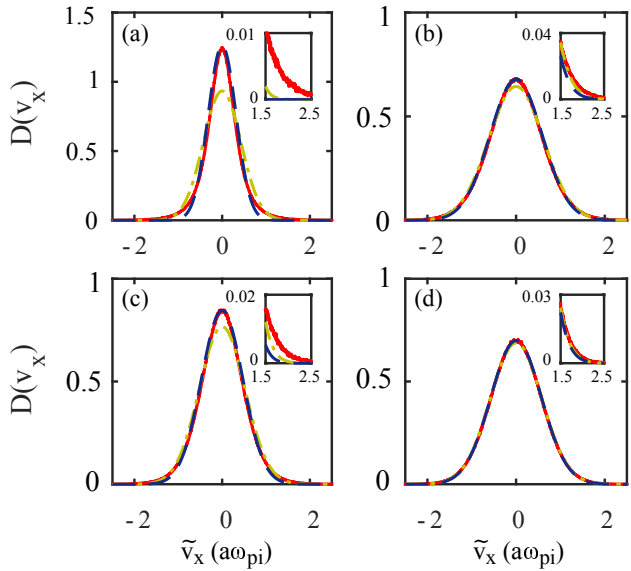}
\caption{One-dimensional velocity distribution for particles in a Yukawa OCP calculated with the MD simulation (red, solid) compared to maxwellian distribution of equivalent energy (gold, dash-dot) and the maxwellian corresponding to the best fit to the distribution (blue, dashed) for $\kappa=0.39$.  (a)-(d) $\omega_{pi}t/2\pi=$\{$0.1,0.25,0.5,3$\}.  Inset: The ``wing'' of the distribution, showing the relatively large populations at high velocity in the distribution calculated with MD.}
\label{fig:distroComparison}
\end{figure}

Figure~\ref{fig:distroComparison} shows the velocity distribution (in normalized units $\tilde{v}=v/(a\omega_{pi})$) from the MD simulation (red) taken at various times throughout the equilibration process.  The distribution is clearly non-maxwellian at early times, as the real distribution differs from that of a thermalized system with equal total kinetic energy (gold), a feature that was demonstrated in previous MD simulations \cite{bergeson:DIH}.  

\begin{figure}[!h]
  \includegraphics{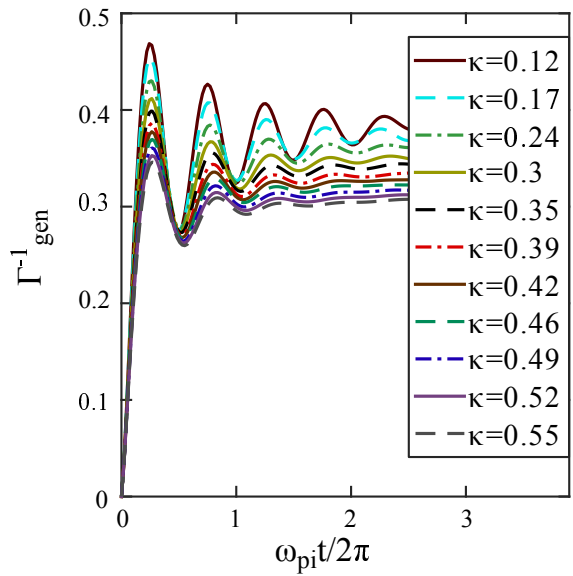}
\caption{Scaled kinetic energy $\Gamma^{-1}_{gen}=2\langle KE\rangle/E_{c}$, where $\langle KE\rangle$ is the one dimensional kinetic energy per particle and $E_{c}=e^{2}/(4\pi\epsilon_{0}a)$ is the Coulomb energy between nearest neighbors, vs scaled time $\omega_{pi}t/2\pi$.  Curves at various $\kappa$ are calculated using a MD simulation that propagates the equations of motion (in scaled units) for a Yukawa OCP with random initial positions and initial velocities all set to zero.}
\label{fig:scaledEnergyVTime}
\end{figure}

The actual kinetic energy per particle at each timestep can be determined by calculating the rms velocity, regardless of whether or not the system is thermalized.  The scaled kinetic energy can then be parameterized by a generalized coupling parameter $\Gamma_{gen}=E_{C}/(2\langle KE\rangle)$, which becomes equal to $\Gamma$ (Eq.~\ref{CCP}) when the plasma is thermalized.  The kinetic energy in natural units is simply $\Gamma_{gen}^{-1}/2$ (we plot $\Gamma_{gen}^{-1}$ instead of $\Gamma_{gen}^{-1}/2$ for convenience).  Figure~\ref{fig:scaledEnergyVTime} shows the DIH curves $\Gamma^{-1}_{gen}$ vs $\omega_{pi}t/2\pi$ for the 10 different values of $\kappa$ for which simulations were conducted (each curve is an average over 100 runs).  In general, the oscillation frequency is roughly $2\omega_{pi}$, and the equilibrium temperature and oscillation amplitude decrease with $\kappa$, while the damping increases with $\kappa$.

\begin{figure}[!h]
  \includegraphics{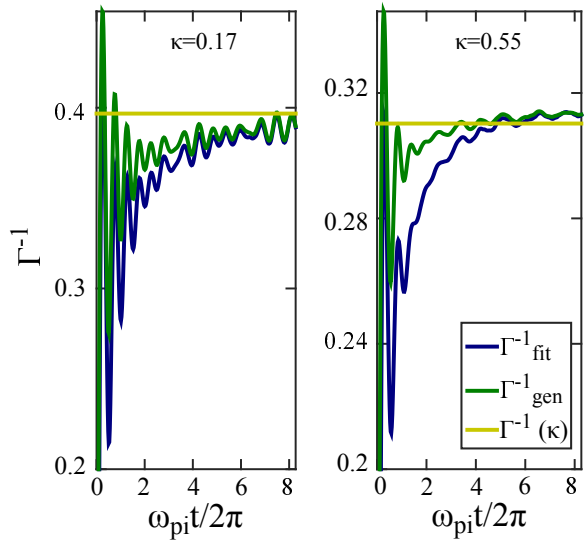}
\caption{Comparison of $\Gamma_{gen}$ and $\Gamma_{fit}$ from MD simulations to $\Gamma(\kappa)$ (Eq.~\ref{DIHGammaSolve}).  $\Gamma_{gen}=E_{C}/(2\langle KE\rangle)$ is the generalized Coulomb coupling parameter calculated from the averaged kinetic energy of the ions, while $\Gamma_{fit}=E_{C}/(k_{B}T_{fit})$ corresponds to the temperature extracted from fitting the velocity distribution to a maxwellian.  In the early stages of the equilibration, $\Gamma_{fit}$ falls below $\Gamma_{gen}$.  As the system equilibrates, both measurements of $\Gamma$ slowly rise to $\Gamma(\kappa)$.}
\label{fig:compToMurillo}
\end{figure}

Experimentally, we do not have access to the complete velocity distribution without some filtering, as it is convolved with a lorentzian by the LIF diagnostic process, as explained in the previous section.  With the signal-to-noise ratio of our experimental data, we cannot unambiguously detect deviations from a maxwellian velocity distribution during the DIH stage (compare the red and blue curves in Fig.~\ref{fig:distroComparison}).  Thus, in order to compare the simulation with the experiment, we run the distribution from the simulation through the same convolution that occurs for our experimental data.  Specifically, we convert the distribution to a doppler-broadened frequency distribution and then numerically convolve it with a lorentzian of width $\gamma$.  We then fit the result to Eq.~\ref{specAssumeGauss} to determine $T_{fit}$, which assumes that the real distribution is maxwellian.  This $T_{fit}$ can then be directly compared to $T_{fit}$ measured in the experiment, or, equivalently, we can convert 
both values of $T_{fit}$ to $\Gamma_{fit}^{-1}$.  $\Gamma_{fit}^{-1}$ tends to slightly underestimate the real kinetic energy of the system (see Fig.~\ref{fig:compToMurillo}) due to the insensitivity of the fit to high velocity ions in the tail of the distribution (Fig.~\ref{fig:distroComparison}).  But, $\Gamma_{fit}$ provides a well-defined prescription for analyzing our numerical and experimental data.  We note that $\Gamma_{fit}$ introduces a dependence on other parameters outside of the Yukawa potential, such as $\sqrt{k_{B}T_{fit}/m_{i}}/(\lambda\gamma)$.  So, $\Gamma_{fit}$ does not rigorously scale with $\kappa$.  However, utilizing the results from the MD simulation, we have confirmed that the effect is not detectable within the current signal-to-noise ratio of the experiment for our parameter range.

We also observe in Fig.~\ref{fig:compToMurillo} that full equilibration to the expected value of $\Gamma$ occurs over a relatively long time scale.  We discuss this further in Sec.~\ref{slowEquilibration}.

\section{Results and Discussion}
\label{Results}

\subsection{Examination of universality of DIH}
\label{sub:universality}

\begin{figure}[!h]
  \includegraphics{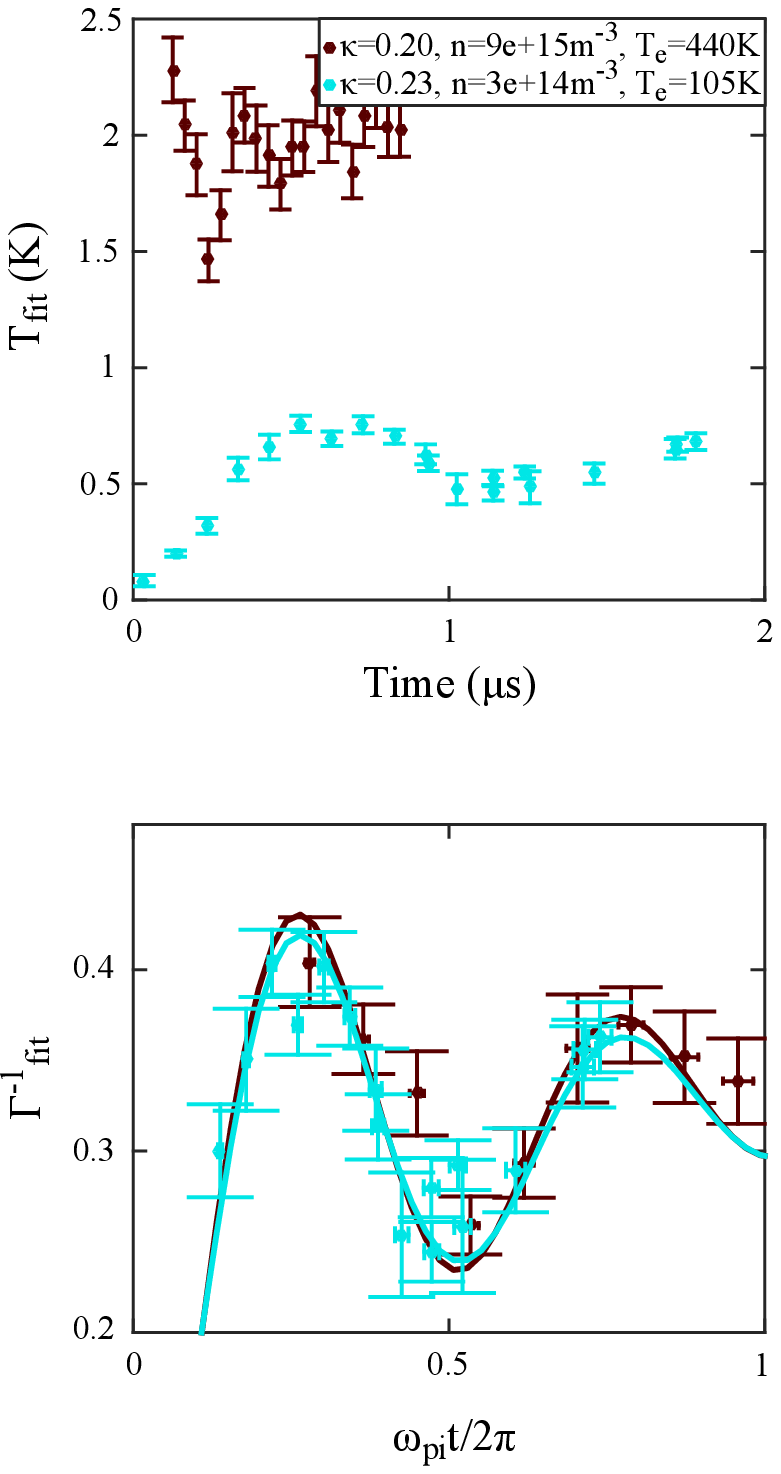}
\caption{Top: $T_{fit}(t)$ for \{$n$,$T_{e}$,$\kappa$\}=\{$3\times10^{14}$m$^{-3}$, $105$\,K, $0.23$\} and \{$9\times10^{15}$m$^{-3}$, $440$\,K, $0.20$\}.  Bottom: $\Gamma^{-1}(\omega_{pi}t/2\pi)$ for the same parameters.  The collapse of the curves in the top panel onto the nearly identical curves matching the MD simulations in the bottom panel demonstrates the universal scaling of DIH over a wide range of $n$ and $T_{e}$.}
\label{fig:DIHSameKapDiffNandTe}
\end{figure}

In order to verify the universal scaling of DIH expected from the Yukawa model, we experimentally measured DIH curves for two conditions with approximately equal $\kappa$: \{$n$,$T_{e}$,$\kappa$\}=\{$3\times10^{14}$m$^{-3}$, $105$\,K, $0.23$\} and \{$9\times10^{15}$m$^{-3}$, $440$\,K, $0.20$\}, where $n$ is measured from the LIF image and $T_{e}$ is set by the wavelength of the ionization laser.  The results are shown in Fig.~\ref{fig:DIHSameKapDiffNandTe}.  Although the two $T_{fit}(t)$ curves differ dramatically, the scaled $\Gamma_{fit}^{-1}(\omega_{pi}t/2\pi)$ curves collapse onto the simulation curves.  This demonstrates the power of universal scaling for Yukawa systems.  Conversely, one could interpret this as experimental evidence that ultracold neutral plasmas are nearly perfect realizations of Yukawa OCPs.

\begin{figure}[!h]
  \includegraphics{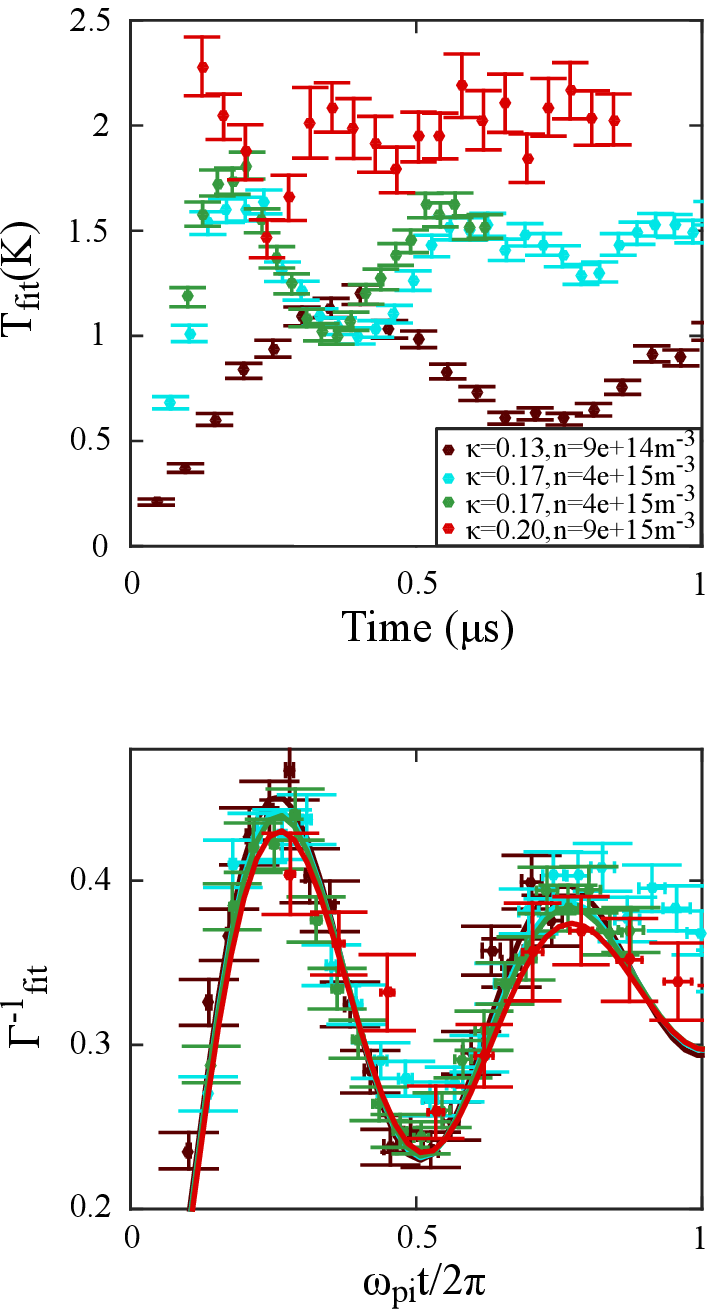}
\caption{Top: $T_{fit}(t)$ for densities ranging from $n=9\times10^{14}$m$^{-3}$ to $n=9\times10^{15}$m$^{-3}$ at $T_{e}=440$\,K.  Bottom: $\Gamma_{fit}^{-1}(\omega_{pi}t/2\pi)$ for the same parameters.  The scaled data is all similar because $\kappa$ varies slowly with plasma density}
\label{fig:DIHDataFano}
\end{figure}

\begin{figure}[!h]
  \includegraphics{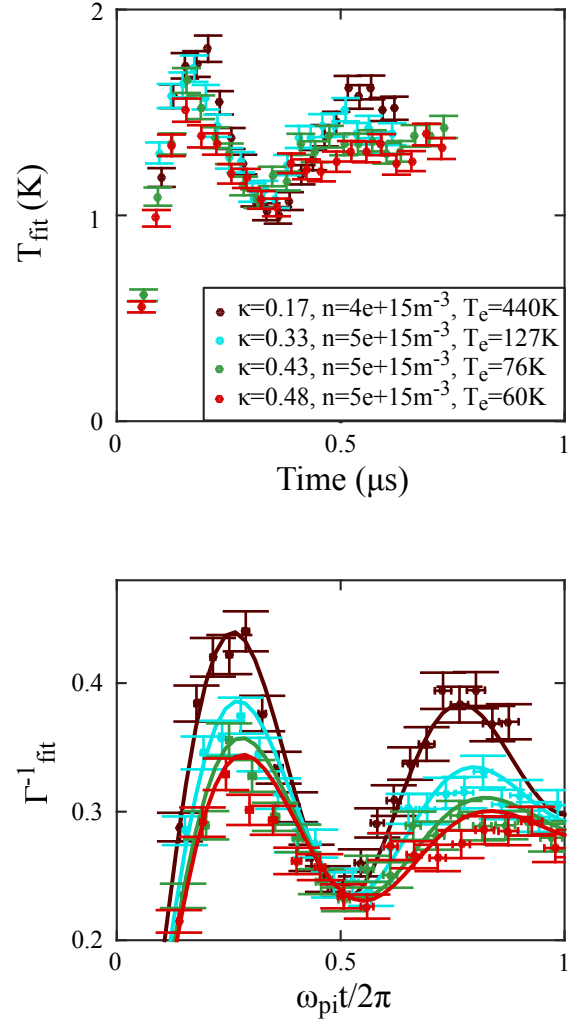}
\caption{Top: $T_{fit}(t)$ for $n\sim5\times10^{15}$m$^{-3}$ at various $\kappa$.  Bottom: $\Gamma_{fit}^{-1}(\omega_{pi}t/2\pi)$ for the same parameters.  Each experimental curve matches the appropriate simulation curve.}
\label{fig:DIHDataVaryKappa}
\end{figure}

Next, we took two sets of data where $\kappa$ was varied; one where we kept $T_{e}$ constant and varied $n$ and another where we kept $n$ relatively constant and varied $T_{e}$.  The results are shown in Figs.~\ref{fig:DIHDataFano} and~\ref{fig:DIHDataVaryKappa}, respectively.  In the constant $T_{e}$ case, we observe that even changing the density by a factor of 10 is not enough to significantly change the behavior of the scaled curves.  This follows from $\kappa$ having a weak dependence on $n$ ($\kappa\propto n^{1/6}$).  In contrast, when $T_{e}$ is varied by a factor of $\sim7$, we see clear differences between the scaled curves as $\kappa\propto T_{e}^{-1/2}$.  Moreover, the experimental results fall on the corresponding MD curves, showing quantitative agreement between data and simulation.

\subsection{Equilibration to $\Gamma_{eq}(\kappa)$}
\label{slowEquilibration}

From the MD simulations (Fig.~\ref{fig:scaledEnergyVTime} and Fig.~\ref{fig:compToMurillo}), it is evident that the average kinetic energy quickly rises to roughly 95\% of the final equilibration value ($\Gamma(\kappa)$, Eq.~\ref{DIHGammaSolve}) within a time of roughly $2\pi/\omega_{pi}$.  However, the final approach to equilibrium as the KEOs damp occurs on a timescale that is an order of magnitude longer.  We confirmed that this behavior is not a numerical artifact by checking convergence of the results with increasing particle number and decreasing time step.  We also note that energy is conserved over the course of the simulation to better than a part in $10^{5}$.

This suggests that equilibration of a Yukawa OCP after a quench of the interaction might display prethermalization dynamics, which has been discussed as a general phenomena of many-body systems far from equilibrium \cite{wetterich:Prethermalization} and recently observed in isolated quantum systems \cite{schmiedmayer:Prethermalization}.  A possible physical explanation for the system studied here might be that the small fraction of high velocity ions observed in the wings of the spectrum (see Fig.~\ref{fig:distroComparison}) takes a long time to reach equilibrium.  This is somewhat expected because the ion-ion collision rate scales with $1/v^{3}$ \cite{spitzer:LSTheory} (although, this scaling is modified slightly in strongly coupled plasmas \cite{Killian:VAF}).

Currently, we cannot confirm this behavior in our experiment.  This is because the timescale is long enough for other temperature-changing dynamics, such as heating from electron-ion collisions and expansion-induced adiabatic cooling \cite{killian:UNPEvolve}, to mask the effect.  

\subsection{Application: Using DIH to measure density and electron temperature}
\label{nAndTeFit}

A typical challenge in UNP experiments is the precise measurement of plasma density and electron temperature.  As discussed and demonstrated in the previous sections, an experimental DIH curve $T_{fit}(t)$ taken at a known $n$ and $T_{e}$ will match the $\Gamma^{-1}_{fit}(\omega_{pi}t,\kappa)$ from MD simulations corresponding to $\kappa(n,T_{e})$ after $T_{fit}$ and $t$ are scaled by $E_{c}/k_{B}$ and $\omega_{pi}^{-1}$ respectively.  Conversely, one could use a fitting routine to match an experimental DIH curve to a simulated $T_{fit}$ curve with density and electron temperature as fit parameters. 

To generate the simulated $T_{fit}$ curve for a given $n$ and $T_{e}$, we utilize a ``library'' of MD results.  MD simulations were conducted at 10 different values of $\kappa$ (see Fig.~\ref{fig:scaledEnergyVTime}).  For each simulation, we record the 1D velocity distribution, $D(\tilde{v})$, every 40 timesteps (0.14$\omega_{pi}t$), where the natural velocity unit $\tilde{v}=v/(a\omega_{pi})$ (see Fig.~\ref{fig:distroComparison}).  The velocity distributions for arbitrary $0.12\le\kappa\le0.55$ determined by the input $n$ and $T_{e}$ can then be determined by interpolating between the 10 acquired sets of distributions.  The simulated $T_{fit}$ curve is then determined by first unscaling the interpolated velocity distributions so that they are in units of m/s, then convolving them with a lorentzian (as in Eq.~\ref{specArbDistro}) and fitting the results to Eq.~\ref{specAssumeGauss}.  Adjusting $n$ and $T_{e}$ to minimize the difference between this simulated $T_{fit}$ curve and the experimental $T_{fit}$ 
data determines best fit values and confidence intervals for both parameters.

\begin{figure}[!h]
  \includegraphics{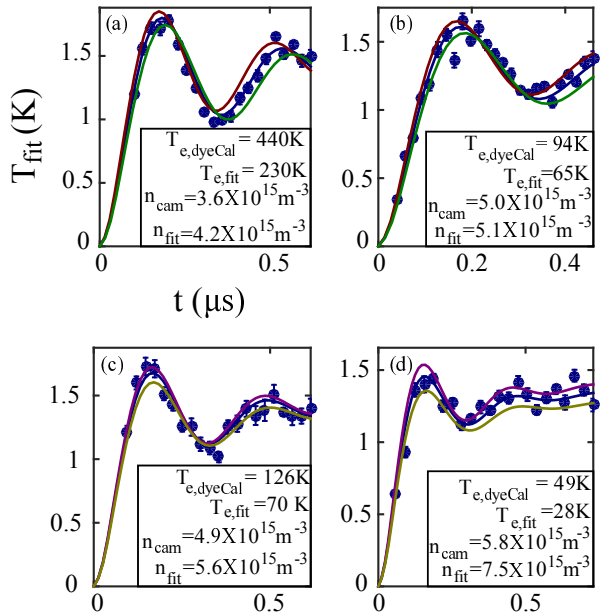}
\caption{Demonstration of effectiveness of fitting DIH curves for $n$ and $T_{e}$.  (a)-(b): Curve of best fit (blue) with curves corresponding to +10\% (red) and -10\% (green) variation in $n_{fit}$.  (c)-(d): Curve of best fit (blue) with curves corresponding to +20\% (purple) and -20\% (gold) variation in $T_{e,fit}$.}
\label{fig:fitEfficacy}
\end{figure}

\begin{figure}[!h]
  \includegraphics{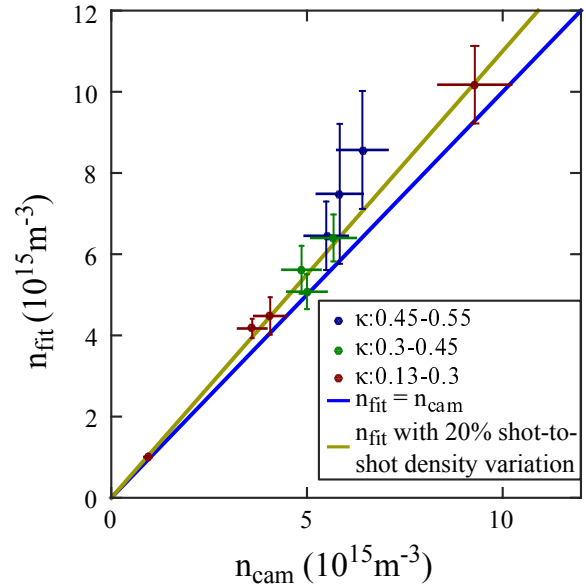}
\caption{Comparison between $n_{cam}$ and $n_{fit}$.  Uncertainty in $n_{cam}$ stems from from camera calibration ($\pm10\%$).  Uncertainty in $n_{fit}$ is determined directly from the confidence intervals (95\% confidence) of the fit.  The gold curve shows the expected $n_{fit}$ for a plasma with additional density variation as described in the text.}
\label{fig:nFitVnCam}
\end{figure}

\begin{figure}[!h]
  \includegraphics{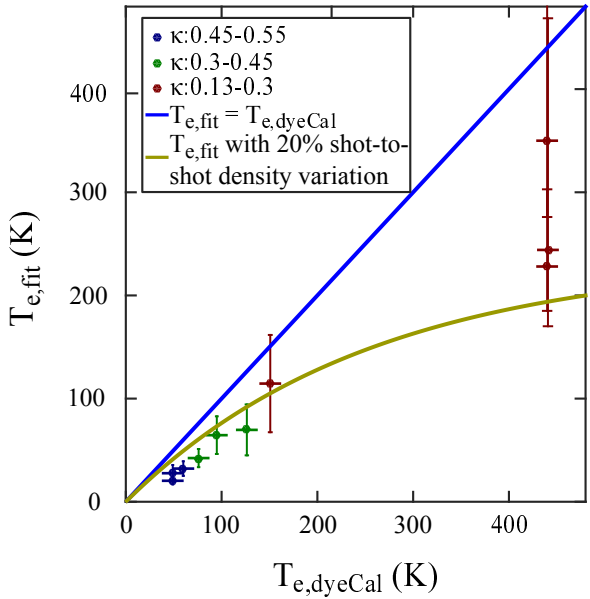}
\caption{$T_{e,fit}$ vs $T_{e,dyeCal}$  Uncertainty in $T_{e,dyeCal}$ stems from uncertainty in the dye laser frequency ($\pm10\,K$).  Uncertainty in $T_{e,fit}$ is determined from the confidence intervals (95\% confidence) of the fit.  The gold curve shows the expected $T_{e,fit}$ for a plasma with additional density variation as described in the text.}
\label{fig:TeFitVTeDial}
\end{figure}

We applied the fitting routine to 10 distinct $T_{fit}(t)$ curves; four are reproduced in Fig.~\ref{fig:fitEfficacy}.  We also compared the fitted density ($n_{fit}$) and electron temperature ($T_{e,fit}$) to the density determined from the LIF measurements ($n_{cam}$) and the electron temperature measured from the wavelength of our pulsed dye laser ($T_{e,dyeCal}$), as illustrated in Figures~\ref{fig:nFitVnCam} and~\ref{fig:TeFitVTeDial}.  The uncertainties in both fit parameters are taken from the 95\% confidence interval of the fit routine.  The uncertainty in $n_{cam}$ is systematically around 20\% due mostly to imprecise knowledge of the shape of the plasma along the axis of the absorption imaging beam, whereas the uncertainty in $T_{e,dyeCal}$ is taken to be $\pm10$\,K.

We observe clear systematic deviations in the fit measurements of both parameters; the fitted densities are too high (Fig.~\ref{fig:nFitVnCam}) and the fitted electron temperatures are too low (Fig.~\ref{fig:TeFitVTeDial}).  These deviations can, at least in part, be explained by additional dephasing in the DIH curve caused by density fluctuations: Fluctuations arise from shot-to-shot experimental variation and the inhomogeneous, gaussian density distribution of the plasma.  The measurement effectively averages oscillations of different frequencies due to the dependence of $\omega_{pi}$ on $n$.  The additional dephasing mimics the damping that would result from increasing the screening parameter $\kappa$, therefore, this dephasing should push $T_{e,fit}$ down with respect to $T_{e,dyeCal}$, as $\kappa\propto T_{e}^{-1/2}$.   However, a higher fitted $\kappa$ value also would reduce the expected DIH for a plasma of density $n_{cam}$ (see Fig.~\ref{fig:GamVKap}); $n_{fit}$ is therefore pushed higher than $n_{
cam}$ in order to increase the energy scaling parameter $E_{c}$ enough to compensate for the reduction in DIH.  The difference between $n_{fit}$ and $n_{cam}$ has a negligible effect on the fitted $\kappa$ relative to the difference between $T_{e,fit}$ and $T_{e,dyeCal}$.  

We simulate this effect using our MD library as discussed in further detail in the Appendix.  By fitting the simulated curves in the same way we fit experimental $T_{fit}(t)$ data, we acquired the gold curves in Figs.~\ref{fig:nFitVnCam} and~\ref{fig:TeFitVTeDial}, which largely account for the observed deviation in both $n$ and $T_{e}$.  These curves correspond to a 20\% shot-to-shot variation in density, which agrees with the estimate of shot-to-shot density variation obtained from images of individual plasmas.  For both parameters, the deviation of the fit value from the actual value varies weakly with $\kappa$.  But, this effect is small compared to the size of the deviation ($<25$\% in our range).  So, we neglect it here, and show a single curve of the fit parameters versus the actual plasma parameters to describe all the $\kappa$ values in our range.

\section{Conclusion}

This work represents a clear demonstration of the universal scaling of Yukawa OCP dynamics with respect to $\kappa$.  We have confirmed that plasma dynamics after a rapid quench from $\kappa_{0}=\infty$ to $\kappa(n,T_{e})$ are universal in $\kappa$ over nearly two orders of magnitude in $n$.  This work expands on prior studies of DIH in UNPs \cite{killian:DIH,bergeson:DIH,bergeson:DIHScreen} by showing with simulation and experiment that the universal scaling holds over the entire equilibration period, encompassing the initial disorder-induced heating phase, damped kinetic energy oscillations, and the slow approach to equilibrium over several inverse plasma periods.  We have also demonstrated that universal scaling can be used as a tool for measuring the plasma density and, to a lesser degree, the electron temperature.  Future work will focus on the excitation and damping of collective modes during equilibration, and the influence of $\kappa$ on these phenomena.

\begin{acknowledgments}
This work was supported by the National Science Foundation (PHY-0714603), Air Force Office of Scientific Research (FA9550-12-1-0267), Department of Energy, Fusion Energy Sciences (DE-SC0014455), Department of Defense (DoD) through the National Defense Science \& Engineering Graduate Fellowship (NDSEG), and Program NIH award NCRR S10RR02950, an IBM Shared University Research (SUR) Award in partnership with CISCO, Qlogic and Adaptive Computing, and Rice University.
\end{acknowledgments}

\appendix*

\section{Effect of Density Variation}
\label{DensityVariation}

Here we consider the effect of experimental density variations on our analysis of DIH and the KEOs.

Density fluctuations arise in two ways in the experiment.  First, underlying each LIF spectrum are images of $\sim1000$ plasmas, which is necessary for good statistics.  Ideally, all of those plasmas would have the same size, density, electron temperature, etc.  However, there is natural shot to shot fluctuation in all of these parameters.  

Second, plasmas inherit the gaussian spatial distribution of the MOT, i.e., they have non-uniform density.  The largest region that is used when measuring $T_{fit}$ is a $1\sigma\times1\sigma\times1\sigma$ region in the center of the plasma, selected by the LIF laser beam and region of interest on the camera.  The variation in density within this region forms another source of density variation.

\begin{figure}[!h]
  \includegraphics{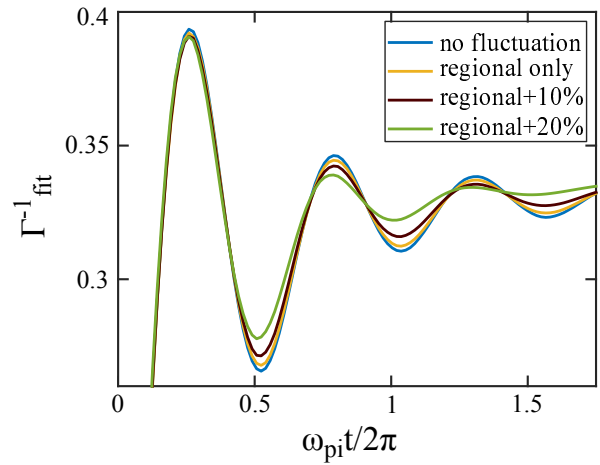}
\caption{Comparison between the uniform density $\kappa=0.35$ MD DIH curve (blue) and the curves with regional and/or shot-to-shot density fluctuations taken into account.  The shot-to-shot fluctuations are the larger source of dephasing, becoming significant when the fluctuations are on the 20\% level (green)}
\label{fig:Dephasing}
\end{figure}

The spread in $\kappa$ and $E_{c}$ from density variations is small.  However, since the frequency of the KEOs depends on density through $\omega_{pi}$, averaging curves corresponding to different densities results in apparent increased damping through dephasing.  In Fig.~\ref{fig:Dephasing}, we examine the effect of both sources of density fluctuations using the MD data.  

For a given peak density $n_{0}$ and temperature $T_{e}$, the regional variation is described by dividing the $1\sigma\times1\sigma\times1\sigma$ box into sub-boxes of size $0.05\sigma\times0.05\sigma\times0.05\sigma$, with the density in each sub-box calculated using the measured gaussian density distribution $n_{s}=n_{0}\exp\left[-\frac{x^{2}+y^{2}+z^{2}}{2\sigma^{2}}\right]$, where $(x,y,z)$ are the sub-box coordinates.  We then unscale MD DIH data for $\kappa(n_{s},T_{e})$ in each sub-box (i.e., in each sub-box we convert the MD $\Gamma^{-1}_{fit}(\omega_{pi}t/2\pi)$ to $T_{fit}(t)$ with the scaling factors determined by $n_{s}$). $T_{fit}(t)$ curves are then averaged together and then rescaled back to $\Gamma_{fit}$ using $n_{0}$.  The shot-to-shot fluctuations are taken into account by repeating that procedure for a set of 1000 values of $n_{0}$ taken from a normal distribution with standard deviation of either 10\% or 20\%.  Figure~\ref{fig:Dephasing} shows the effect of density 
variation.  The regional variation of density has a small effect on the fit parameters, but the shot-to-shot fluctuations are significant.  By fitting this simulated data for $n$ and $T_{e}$ in the same way we fit our real experimental data, we determine that the fitted electron temperature is reduced, as the additional dephasing is indistinguishable from a increase in $\kappa$, and thus a decrease in $T_{e}$ (see Fig.~\ref{fig:TeFitVTeDial}).  The density measurement is also increased (Fig.~\ref{fig:nFitVnCam}).


\begin{thebibliography}{36}%
\makeatletter
\providecommand \@ifxundefined [1]{%
 \@ifx{#1\undefined}
}%
\providecommand \@ifnum [1]{%
 \ifnum #1\expandafter \@firstoftwo
 \else \expandafter \@secondoftwo
 \fi
}%
\providecommand \@ifx [1]{%
 \ifx #1\expandafter \@firstoftwo
 \else \expandafter \@secondoftwo
 \fi
}%
\providecommand \natexlab [1]{#1}%
\providecommand \enquote  [1]{``#1''}%
\providecommand \bibnamefont  [1]{#1}%
\providecommand \bibfnamefont [1]{#1}%
\providecommand \citenamefont [1]{#1}%
\providecommand \href@noop [0]{\@secondoftwo}%
\providecommand \href [0]{\begingroup \@sanitize@url \@href}%
\providecommand \@href[1]{\@@startlink{#1}\@@href}%
\providecommand \@@href[1]{\endgroup#1\@@endlink}%
\providecommand \@sanitize@url [0]{\catcode `\\12\catcode `\$12\catcode
  `\&12\catcode `\#12\catcode `\^12\catcode `\_12\catcode `\%12\relax}%
\providecommand \@@startlink[1]{}%
\providecommand \@@endlink[0]{}%
\providecommand \url  [0]{\begingroup\@sanitize@url \@url }%
\providecommand \@url [1]{\endgroup\@href {#1}{\urlprefix }}%
\providecommand \urlprefix  [0]{URL }%
\providecommand \Eprint [0]{\href }%
\providecommand \doibase [0]{http://dx.doi.org/}%
\providecommand \selectlanguage [0]{\@gobble}%
\providecommand \bibinfo  [0]{\@secondoftwo}%
\providecommand \bibfield  [0]{\@secondoftwo}%
\providecommand \translation [1]{[#1]}%
\providecommand \BibitemOpen [0]{}%
\providecommand \bibitemStop [0]{}%
\providecommand \bibitemNoStop [0]{.\EOS\space}%
\providecommand \EOS [0]{\spacefactor3000\relax}%
\providecommand \BibitemShut  [1]{\csname bibitem#1\endcsname}%
\let\auto@bib@innerbib\@empty
%</preamble>
\bibitem [{\citenamefont {Salpeter}\ and\ \citenamefont
  {Horn}(1969)}]{salpeter:WhiteDwarf}%
  \BibitemOpen
  \bibfield  {author} {\bibinfo {author} {\bibfnamefont {E.}~\bibnamefont
  {Salpeter}}\ and\ \bibinfo {author} {\bibfnamefont {H.~V.}\ \bibnamefont
  {Horn}},\ }\href@noop {} {\bibfield  {journal} {\bibinfo  {journal}
  {Astrophysical Journal}\ }\textbf {\bibinfo {volume} {155}},\ \bibinfo
  {pages} {183} (\bibinfo {year} {1969})}\BibitemShut {NoStop}%
\bibitem [{\citenamefont {Stevenson}(1980)}]{stevenson:Jovian}%
  \BibitemOpen
  \bibfield  {author} {\bibinfo {author} {\bibfnamefont {D.}~\bibnamefont
  {Stevenson}},\ }\href@noop {} {\bibfield  {journal} {\bibinfo  {journal} {J.
  Phys. Colloques}\ }\textbf {\bibinfo {volume} {41}},\ \bibinfo {pages} {53}
  (\bibinfo {year} {1980})}\BibitemShut {NoStop}%
\bibitem [{\citenamefont {Remington}\ \emph {et~al.}(2008)\citenamefont
  {Remington}, \citenamefont {Drake},\ and\ \citenamefont
  {Ryutov}}]{remington:Astro}%
  \BibitemOpen
  \bibfield  {author} {\bibinfo {author} {\bibfnamefont {B.}~\bibnamefont
  {Remington}}, \bibinfo {author} {\bibfnamefont {R.}~\bibnamefont {Drake}}, \
  and\ \bibinfo {author} {\bibfnamefont {D.}~\bibnamefont {Ryutov}},\
  }\href@noop {} {\bibfield  {journal} {\bibinfo  {journal} {Rev. Mod. Phys}\
  }\textbf {\bibinfo {volume} {78}},\ \bibinfo {pages} {755} (\bibinfo {year}
  {2008})}\BibitemShut {NoStop}%
\bibitem [{\citenamefont {Lindl}(1995)}]{lindl:ICF}%
  \BibitemOpen
  \bibfield  {author} {\bibinfo {author} {\bibfnamefont {J.}~\bibnamefont
  {Lindl}},\ }\href@noop {} {\bibfield  {journal} {\bibinfo  {journal} {Physics
  of Plasmas}\ }\textbf {\bibinfo {volume} {2}},\ \bibinfo {pages} {3933}
  (\bibinfo {year} {1995})}\BibitemShut {NoStop}%
\bibitem [{\citenamefont {Schella}\ \emph {et~al.}(2014)\citenamefont
  {Schella}, \citenamefont {Mulsow},\ and\ \citenamefont
  {Melzer}}]{Melzer:CorrReCryst}%
  \BibitemOpen
  \bibfield  {author} {\bibinfo {author} {\bibfnamefont {A.}~\bibnamefont
  {Schella}}, \bibinfo {author} {\bibfnamefont {M.}~\bibnamefont {Mulsow}}, \
  and\ \bibinfo {author} {\bibfnamefont {A.}~\bibnamefont {Melzer}},\
  }\href@noop {} {\bibfield  {journal} {\bibinfo  {journal} {Physics of
  Plasmas}\ }\textbf {\bibinfo {volume} {21}},\ \bibinfo {pages} {050701}
  (\bibinfo {year} {2014})}\BibitemShut {NoStop}%
\bibitem [{\citenamefont {Morfill}\ and\ \citenamefont
  {Ivlev}(2009)}]{morfill:Dusty}%
  \BibitemOpen
  \bibfield  {author} {\bibinfo {author} {\bibfnamefont {G.}~\bibnamefont
  {Morfill}}\ and\ \bibinfo {author} {\bibfnamefont {A.}~\bibnamefont
  {Ivlev}},\ }\href@noop {} {\bibfield  {journal} {\bibinfo  {journal} {Rev.
  Mod. Phys}\ }\textbf {\bibinfo {volume} {81}},\ \bibinfo {pages} {1353}
  (\bibinfo {year} {2009})}\BibitemShut {NoStop}%
\bibitem [{\citenamefont {Williams}\ \emph {et~al.}(1976)\citenamefont
  {Williams}, \citenamefont {Crandall},\ and\ \citenamefont
  {Wojtowicz}}]{wojtowicz:ColloidalMelting}%
  \BibitemOpen
  \bibfield  {author} {\bibinfo {author} {\bibfnamefont {R.}~\bibnamefont
  {Williams}}, \bibinfo {author} {\bibfnamefont {R.}~\bibnamefont {Crandall}},
  \ and\ \bibinfo {author} {\bibfnamefont {P.}~\bibnamefont {Wojtowicz}},\
  }\href@noop {} {\bibfield  {journal} {\bibinfo  {journal} {Phys. Rev. Lett.}\
  }\textbf {\bibinfo {volume} {37}},\ \bibinfo {pages} {348} (\bibinfo {year}
  {1976})}\BibitemShut {NoStop}%
\bibitem [{\citenamefont {Kremer}\ \emph {et~al.}(1986)\citenamefont {Kremer},
  \citenamefont {Robbins},\ and\ \citenamefont
  {Grest}}]{grest:YukawaModelColloids}%
  \BibitemOpen
  \bibfield  {author} {\bibinfo {author} {\bibfnamefont {K.}~\bibnamefont
  {Kremer}}, \bibinfo {author} {\bibfnamefont {M.}~\bibnamefont {Robbins}}, \
  and\ \bibinfo {author} {\bibfnamefont {G.}~\bibnamefont {Grest}},\
  }\href@noop {} {\bibfield  {journal} {\bibinfo  {journal} {Phys. Rev. Lett.}\
  }\textbf {\bibinfo {volume} {57}},\ \bibinfo {pages} {2694} (\bibinfo {year}
  {1986})}\BibitemShut {NoStop}%
\bibitem [{\citenamefont {Simien}\ \emph {et~al.}(2004)\citenamefont {Simien},
  \citenamefont {Chen}, \citenamefont {Gupta}, \citenamefont {Laha},
  \citenamefont {Martinez}, \citenamefont {Mickelson}, \citenamefont {Nagel},\
  and\ \citenamefont {Killian}}]{killian:UNPCreate}%
  \BibitemOpen
  \bibfield  {author} {\bibinfo {author} {\bibfnamefont {C.~E.}\ \bibnamefont
  {Simien}}, \bibinfo {author} {\bibfnamefont {Y.~C.}\ \bibnamefont {Chen}},
  \bibinfo {author} {\bibfnamefont {P.}~\bibnamefont {Gupta}}, \bibinfo
  {author} {\bibfnamefont {S.}~\bibnamefont {Laha}}, \bibinfo {author}
  {\bibfnamefont {Y.~N.}\ \bibnamefont {Martinez}}, \bibinfo {author}
  {\bibfnamefont {P.~G.}\ \bibnamefont {Mickelson}}, \bibinfo {author}
  {\bibfnamefont {S.~B.}\ \bibnamefont {Nagel}}, \ and\ \bibinfo {author}
  {\bibfnamefont {T.~C.}\ \bibnamefont {Killian}},\ }\href@noop {} {\bibfield
  {journal} {\bibinfo  {journal} {Phys. Rev. Lett.}\ }\textbf {\bibinfo
  {volume} {92}},\ \bibinfo {pages} {143001} (\bibinfo {year}
  {2004})}\BibitemShut {NoStop}%
\bibitem [{\citenamefont {Killian}\ \emph {et~al.}(1999)\citenamefont
  {Killian}, \citenamefont {Kulin}, \citenamefont {Bergeson}, \citenamefont
  {Orozco}, \citenamefont {Orzel},\ and\ \citenamefont
  {Rolston}}]{rolston:UNPCreate}%
  \BibitemOpen
  \bibfield  {author} {\bibinfo {author} {\bibfnamefont {T.~C.}\ \bibnamefont
  {Killian}}, \bibinfo {author} {\bibfnamefont {S.}~\bibnamefont {Kulin}},
  \bibinfo {author} {\bibfnamefont {S.~D.}\ \bibnamefont {Bergeson}}, \bibinfo
  {author} {\bibfnamefont {L.~A.}\ \bibnamefont {Orozco}}, \bibinfo {author}
  {\bibfnamefont {C.}~\bibnamefont {Orzel}}, \ and\ \bibinfo {author}
  {\bibfnamefont {S.~L.}\ \bibnamefont {Rolston}},\ }\href@noop {} {\bibfield
  {journal} {\bibinfo  {journal} {Phys. Rev. Lett.}\ }\textbf {\bibinfo
  {volume} {83}},\ \bibinfo {pages} {4776} (\bibinfo {year}
  {1999})}\BibitemShut {NoStop}%
\bibitem [{\citenamefont {Hamaguchi}\ \emph {et~al.}(1997)\citenamefont
  {Hamaguchi}, \citenamefont {Farouki},\ and\ \citenamefont
  {Dubin}}]{Hamaguchi:UCalc}%
  \BibitemOpen
  \bibfield  {author} {\bibinfo {author} {\bibfnamefont {S.}~\bibnamefont
  {Hamaguchi}}, \bibinfo {author} {\bibfnamefont {R.~T.}\ \bibnamefont
  {Farouki}}, \ and\ \bibinfo {author} {\bibfnamefont {D.~H.~E.}\ \bibnamefont
  {Dubin}},\ }\href@noop {} {\bibfield  {journal} {\bibinfo  {journal} {Phys.
  Rev. E.}\ }\textbf {\bibinfo {volume} {56}},\ \bibinfo {pages} {4671}
  (\bibinfo {year} {1997})}\BibitemShut {NoStop}%
\bibitem [{\citenamefont {Farouki}\ and\ \citenamefont
  {Hamaguchi}(1994)}]{hamaguchi:YukawaMD}%
  \BibitemOpen
  \bibfield  {author} {\bibinfo {author} {\bibfnamefont {R.}~\bibnamefont
  {Farouki}}\ and\ \bibinfo {author} {\bibfnamefont {S.}~\bibnamefont
  {Hamaguchi}},\ }\href@noop {} {\bibfield  {journal} {\bibinfo  {journal} {J.
  Chem. Phys.}\ }\textbf {\bibinfo {volume} {101}},\ \bibinfo {pages} {9885}
  (\bibinfo {year} {1994})}\BibitemShut {NoStop}%
\bibitem [{\citenamefont {Murillo}(2006)}]{murillo:UFast}%
  \BibitemOpen
  \bibfield  {author} {\bibinfo {author} {\bibfnamefont {M.}~\bibnamefont
  {Murillo}},\ }\href@noop {} {\bibfield  {journal} {\bibinfo  {journal} {Phys.
  Rev. Lett.}\ }\textbf {\bibinfo {volume} {96}},\ \bibinfo {pages} {165001}
  (\bibinfo {year} {2006})}\BibitemShut {NoStop}%
\bibitem [{\citenamefont {Morrison}\ \emph {et~al.}(2012)\citenamefont
  {Morrison}, \citenamefont {Saquet},\ and\ \citenamefont
  {Grant}}]{grant:TBRUnivScaling}%
  \BibitemOpen
  \bibfield  {author} {\bibinfo {author} {\bibfnamefont {J.}~\bibnamefont
  {Morrison}}, \bibinfo {author} {\bibfnamefont {N.}~\bibnamefont {Saquet}}, \
  and\ \bibinfo {author} {\bibfnamefont {E.}~\bibnamefont {Grant}},\
  }\href@noop {} {\bibfield  {journal} {\bibinfo  {journal} {J. Phys. B}\
  }\textbf {\bibinfo {volume} {45}},\ \bibinfo {pages} {025701} (\bibinfo
  {year} {2012})}\BibitemShut {NoStop}%
\bibitem [{\citenamefont {Gupta}\ \emph {et~al.}(2007)\citenamefont {Gupta},
  \citenamefont {Laha}, \citenamefont {Simien}, \citenamefont {Gao},
  \citenamefont {Castro}, \citenamefont {Killian},\ and\ \citenamefont
  {Pohl}}]{killian:electronTemp}%
  \BibitemOpen
  \bibfield  {author} {\bibinfo {author} {\bibfnamefont {P.}~\bibnamefont
  {Gupta}}, \bibinfo {author} {\bibfnamefont {S.}~\bibnamefont {Laha}},
  \bibinfo {author} {\bibfnamefont {C.~E.}\ \bibnamefont {Simien}}, \bibinfo
  {author} {\bibfnamefont {H.}~\bibnamefont {Gao}}, \bibinfo {author}
  {\bibfnamefont {J.}~\bibnamefont {Castro}}, \bibinfo {author} {\bibfnamefont
  {T.~C.}\ \bibnamefont {Killian}}, \ and\ \bibinfo {author} {\bibfnamefont
  {T.}~\bibnamefont {Pohl}},\ }\href@noop {} {\bibfield  {journal} {\bibinfo
  {journal} {Phys. Rev. Lett.}\ }\textbf {\bibinfo {volume} {99}},\ \bibinfo
  {pages} {075005} (\bibinfo {year} {2007})}\BibitemShut {NoStop}%
\bibitem [{\citenamefont {Ichimaru}(1982)}]{ichimaru:SCPReview}%
  \BibitemOpen
  \bibfield  {author} {\bibinfo {author} {\bibfnamefont {S.}~\bibnamefont
  {Ichimaru}},\ }\href@noop {} {\bibfield  {journal} {\bibinfo  {journal} {Rev.
  Mod. Phys}\ }\textbf {\bibinfo {volume} {54}},\ \bibinfo {pages} {1017}
  (\bibinfo {year} {1982})}\BibitemShut {NoStop}%
\bibitem [{\citenamefont {Morrison}\ \emph {et~al.}(2008)\citenamefont
  {Morrison}, \citenamefont {Rennick}, \citenamefont {Keller},\ and\
  \citenamefont {Grant}}]{grant:UNPCreate}%
  \BibitemOpen
  \bibfield  {author} {\bibinfo {author} {\bibfnamefont {J.}~\bibnamefont
  {Morrison}}, \bibinfo {author} {\bibfnamefont {C.}~\bibnamefont {Rennick}},
  \bibinfo {author} {\bibfnamefont {J.}~\bibnamefont {Keller}}, \ and\ \bibinfo
  {author} {\bibfnamefont {E.}~\bibnamefont {Grant}},\ }\href@noop {}
  {\bibfield  {journal} {\bibinfo  {journal} {Phys. Rev. Lett.}\ }\textbf
  {\bibinfo {volume} {101}},\ \bibinfo {pages} {205005} (\bibinfo {year}
  {2008})}\BibitemShut {NoStop}%
\bibitem [{\citenamefont {{Rolston}}\ \emph {et~al.}(1998)\citenamefont
  {{Rolston}}, \citenamefont {{Bergeson}}, \citenamefont {{Kulin}},\ and\
  \citenamefont {{Orzel}}}]{rolston:SpontIonization}%
  \BibitemOpen
  \bibfield  {author} {\bibinfo {author} {\bibfnamefont {S.~L.}\ \bibnamefont
  {{Rolston}}}, \bibinfo {author} {\bibfnamefont {S.~D.}\ \bibnamefont
  {{Bergeson}}}, \bibinfo {author} {\bibfnamefont {S.}~\bibnamefont {{Kulin}}},
  \ and\ \bibinfo {author} {\bibfnamefont {C.}~\bibnamefont {{Orzel}}},\ }in\
  \href@noop {} {\emph {\bibinfo {booktitle} {APS Division of Atomic, Molecular
  and Optical Physics Meeting Abstracts}}}\ (\bibinfo {year} {1998})\ p.\
  \bibinfo {pages} {609}\BibitemShut {NoStop}%
\bibitem [{\citenamefont {Robinson}\ \emph {et~al.}(2000)\citenamefont
  {Robinson}, \citenamefont {Tolra}, \citenamefont {Noel}, \citenamefont
  {Gallagher},\ and\ \citenamefont {Pillet}}]{gallagher:SpontIonization}%
  \BibitemOpen
  \bibfield  {author} {\bibinfo {author} {\bibfnamefont {M.~P.}\ \bibnamefont
  {Robinson}}, \bibinfo {author} {\bibfnamefont {B.~L.}\ \bibnamefont {Tolra}},
  \bibinfo {author} {\bibfnamefont {M.~W.}\ \bibnamefont {Noel}}, \bibinfo
  {author} {\bibfnamefont {T.~F.}\ \bibnamefont {Gallagher}}, \ and\ \bibinfo
  {author} {\bibfnamefont {P.}~\bibnamefont {Pillet}},\ }\href@noop {}
  {\bibfield  {journal} {\bibinfo  {journal} {Phys. Rev. Lett.}\ }\textbf
  {\bibinfo {volume} {85}},\ \bibinfo {pages} {4466} (\bibinfo {year}
  {2000})}\BibitemShut {NoStop}%
\bibitem [{\citenamefont {de~Saint-Vincent}\ \emph {et~al.}(2013)\citenamefont
  {de~Saint-Vincent}, \citenamefont {Hofmann}, \citenamefont {Schempp},
  \citenamefont {Gunter}, \citenamefont {Whitlock},\ and\ \citenamefont
  {Weidemuller}}]{weidemuller:SpontIonization}%
  \BibitemOpen
  \bibfield  {author} {\bibinfo {author} {\bibfnamefont {M.~R.}\ \bibnamefont
  {de~Saint-Vincent}}, \bibinfo {author} {\bibfnamefont {C.~S.}\ \bibnamefont
  {Hofmann}}, \bibinfo {author} {\bibfnamefont {H.}~\bibnamefont {Schempp}},
  \bibinfo {author} {\bibfnamefont {G.}~\bibnamefont {Gunter}}, \bibinfo
  {author} {\bibfnamefont {S.}~\bibnamefont {Whitlock}}, \ and\ \bibinfo
  {author} {\bibfnamefont {M.}~\bibnamefont {Weidemuller}},\ }\href@noop {}
  {\bibfield  {journal} {\bibinfo  {journal} {Phys. Rev. Lett.}\ }\textbf
  {\bibinfo {volume} {110}},\ \bibinfo {pages} {045004} (\bibinfo {year}
  {2013})}\BibitemShut {NoStop}%
\bibitem [{\citenamefont {McQuillen}\ \emph {et~al.}(2015)\citenamefont
  {McQuillen}, \citenamefont {Strickler}, \citenamefont {Langin},\ and\
  \citenamefont {Killian}}]{killian:UNPEvolve}%
  \BibitemOpen
  \bibfield  {author} {\bibinfo {author} {\bibfnamefont {P.}~\bibnamefont
  {McQuillen}}, \bibinfo {author} {\bibfnamefont {T.}~\bibnamefont
  {Strickler}}, \bibinfo {author} {\bibfnamefont {T.}~\bibnamefont {Langin}}, \
  and\ \bibinfo {author} {\bibfnamefont {T.~C.}\ \bibnamefont {Killian}},\
  }\href@noop {} {\bibfield  {journal} {\bibinfo  {journal} {Physics of
  Plasmas}\ }\textbf {\bibinfo {volume} {22}},\ \bibinfo {pages} {033513}
  (\bibinfo {year} {2015})}\BibitemShut {NoStop}%
\bibitem [{\citenamefont {Chen}\ \emph {et~al.}(2004)\citenamefont {Chen},
  \citenamefont {Simien}, \citenamefont {Laha}, \citenamefont {Gupta},
  \citenamefont {Martinez}, \citenamefont {Mickelson}, \citenamefont {Nagel},\
  and\ \citenamefont {Killian}}]{killian:DIH}%
  \BibitemOpen
  \bibfield  {author} {\bibinfo {author} {\bibfnamefont {Y.~C.}\ \bibnamefont
  {Chen}}, \bibinfo {author} {\bibfnamefont {C.~E.}\ \bibnamefont {Simien}},
  \bibinfo {author} {\bibfnamefont {S.}~\bibnamefont {Laha}}, \bibinfo {author}
  {\bibfnamefont {P.}~\bibnamefont {Gupta}}, \bibinfo {author} {\bibfnamefont
  {Y.~N.}\ \bibnamefont {Martinez}}, \bibinfo {author} {\bibfnamefont {P.~G.}\
  \bibnamefont {Mickelson}}, \bibinfo {author} {\bibfnamefont {S.~B.}\
  \bibnamefont {Nagel}}, \ and\ \bibinfo {author} {\bibfnamefont {T.~C.}\
  \bibnamefont {Killian}},\ }\href@noop {} {\bibfield  {journal} {\bibinfo
  {journal} {Phys. Rev. Lett.}\ }\textbf {\bibinfo {volume} {93}},\ \bibinfo
  {pages} {265003} (\bibinfo {year} {2004})}\BibitemShut {NoStop}%
\bibitem [{\citenamefont {Bergeson}\ \emph {et~al.}(2011)\citenamefont
  {Bergeson}, \citenamefont {Denning}, \citenamefont {Lyon},\ and\
  \citenamefont {Robicheaux}}]{bergeson:DIH}%
  \BibitemOpen
  \bibfield  {author} {\bibinfo {author} {\bibfnamefont {S.~D.}\ \bibnamefont
  {Bergeson}}, \bibinfo {author} {\bibfnamefont {A.}~\bibnamefont {Denning}},
  \bibinfo {author} {\bibfnamefont {M.}~\bibnamefont {Lyon}}, \ and\ \bibinfo
  {author} {\bibfnamefont {F.}~\bibnamefont {Robicheaux}},\ }\href@noop {}
  {\bibfield  {journal} {\bibinfo  {journal} {Phys. Rev. A.}\ }\textbf
  {\bibinfo {volume} {83}},\ \bibinfo {pages} {023409} (\bibinfo {year}
  {2011})}\BibitemShut {NoStop}%
\bibitem [{\citenamefont {Lyon}\ \emph {et~al.}(2013)\citenamefont {Lyon},
  \citenamefont {Bergeson},\ and\ \citenamefont
  {Murillo}}]{bergeson:DIHScreen}%
  \BibitemOpen
  \bibfield  {author} {\bibinfo {author} {\bibfnamefont {M.}~\bibnamefont
  {Lyon}}, \bibinfo {author} {\bibfnamefont {S.~D.}\ \bibnamefont {Bergeson}},
  \ and\ \bibinfo {author} {\bibfnamefont {M.}~\bibnamefont {Murillo}},\
  }\href@noop {} {\bibfield  {journal} {\bibinfo  {journal} {Phys. Rev. E.}\
  }\textbf {\bibinfo {volume} {87}},\ \bibinfo {pages} {033101} (\bibinfo
  {year} {2013})}\BibitemShut {NoStop}%
\bibitem [{\citenamefont {Murillo}(2001)}]{murillo:DIHPRL}%
  \BibitemOpen
  \bibfield  {author} {\bibinfo {author} {\bibfnamefont {M.}~\bibnamefont
  {Murillo}},\ }\href@noop {} {\bibfield  {journal} {\bibinfo  {journal} {Phys.
  Rev. Lett.}\ }\textbf {\bibinfo {volume} {87}},\ \bibinfo {pages} {115003}
  (\bibinfo {year} {2001})}\BibitemShut {NoStop}%
\bibitem [{\citenamefont {Gericke}\ and\ \citenamefont
  {Murillo}(2003)}]{murillo:DIH}%
  \BibitemOpen
  \bibfield  {author} {\bibinfo {author} {\bibfnamefont {D.}~\bibnamefont
  {Gericke}}\ and\ \bibinfo {author} {\bibfnamefont {M.}~\bibnamefont
  {Murillo}},\ }\href@noop {} {\bibfield  {journal} {\bibinfo  {journal}
  {Contrib. Plasma Phys}\ }\textbf {\bibinfo {volume} {43}},\ \bibinfo {pages}
  {298} (\bibinfo {year} {2003})}\BibitemShut {NoStop}%
\bibitem [{\citenamefont {Ohta}\ and\ \citenamefont
  {Hamaguchi}(2000)}]{Hamaguchi:VAFOsc}%
  \BibitemOpen
  \bibfield  {author} {\bibinfo {author} {\bibfnamefont {H.}~\bibnamefont
  {Ohta}}\ and\ \bibinfo {author} {\bibfnamefont {S.}~\bibnamefont
  {Hamaguchi}},\ }\href@noop {} {\bibfield  {journal} {\bibinfo  {journal}
  {Physics of Plasmas}\ }\textbf {\bibinfo {volume} {7}},\ \bibinfo {pages}
  {4506} (\bibinfo {year} {2000})}\BibitemShut {NoStop}%
\bibitem [{\citenamefont {Bannasch}\ \emph {et~al.}(2012)\citenamefont
  {Bannasch}, \citenamefont {Castro}, \citenamefont {McQuillen}, \citenamefont
  {Pohl},\ and\ \citenamefont {Killian}}]{Killian:VAF}%
  \BibitemOpen
  \bibfield  {author} {\bibinfo {author} {\bibfnamefont {G.}~\bibnamefont
  {Bannasch}}, \bibinfo {author} {\bibfnamefont {J.}~\bibnamefont {Castro}},
  \bibinfo {author} {\bibfnamefont {P.}~\bibnamefont {McQuillen}}, \bibinfo
  {author} {\bibfnamefont {T.}~\bibnamefont {Pohl}}, \ and\ \bibinfo {author}
  {\bibfnamefont {T.~C.}\ \bibnamefont {Killian}},\ }\href@noop {} {\bibfield
  {journal} {\bibinfo  {journal} {Phys. Rev. Lett.}\ }\textbf {\bibinfo
  {volume} {109}},\ \bibinfo {pages} {185008} (\bibinfo {year}
  {2012})}\BibitemShut {NoStop}%
\bibitem [{\citenamefont {Hansen}\ \emph {et~al.}(1974)\citenamefont {Hansen},
  \citenamefont {Pollock},\ and\ \citenamefont {McDonald}}]{Hansen:NonEqRelax}%
  \BibitemOpen
  \bibfield  {author} {\bibinfo {author} {\bibfnamefont {J.}~\bibnamefont
  {Hansen}}, \bibinfo {author} {\bibfnamefont {E.}~\bibnamefont {Pollock}}, \
  and\ \bibinfo {author} {\bibfnamefont {I.}~\bibnamefont {McDonald}},\
  }\href@noop {} {\bibfield  {journal} {\bibinfo  {journal} {Phys. Rev. Lett.}\
  }\textbf {\bibinfo {volume} {32}},\ \bibinfo {pages} {277} (\bibinfo {year}
  {1974})}\BibitemShut {NoStop}%
\bibitem [{\citenamefont {Gould}\ and\ \citenamefont
  {Mazenko}(1975)}]{Mazenko:NonEqRelax}%
  \BibitemOpen
  \bibfield  {author} {\bibinfo {author} {\bibfnamefont {H.}~\bibnamefont
  {Gould}}\ and\ \bibinfo {author} {\bibfnamefont {G.}~\bibnamefont
  {Mazenko}},\ }\href@noop {} {\bibfield  {journal} {\bibinfo  {journal} {Phys.
  Rev. Lett.}\ }\textbf {\bibinfo {volume} {35}},\ \bibinfo {pages} {1455}
  (\bibinfo {year} {1975})}\BibitemShut {NoStop}%
\bibitem [{\citenamefont {Laha}\ \emph {et~al.}(2007)\citenamefont {Laha},
  \citenamefont {Gupta}, \citenamefont {Simien}, \citenamefont {Gao},
  \citenamefont {Castro}, \citenamefont {Pohl},\ and\ \citenamefont
  {Killian}}]{Killian:LIF}%
  \BibitemOpen
  \bibfield  {author} {\bibinfo {author} {\bibfnamefont {S.}~\bibnamefont
  {Laha}}, \bibinfo {author} {\bibfnamefont {P.}~\bibnamefont {Gupta}},
  \bibinfo {author} {\bibfnamefont {C.~E.}\ \bibnamefont {Simien}}, \bibinfo
  {author} {\bibfnamefont {H.}~\bibnamefont {Gao}}, \bibinfo {author}
  {\bibfnamefont {J.}~\bibnamefont {Castro}}, \bibinfo {author} {\bibfnamefont
  {T.}~\bibnamefont {Pohl}}, \ and\ \bibinfo {author} {\bibfnamefont {T.~C.}\
  \bibnamefont {Killian}},\ }\href@noop {} {\bibfield  {journal} {\bibinfo
  {journal} {Phys. Rev. Lett.}\ }\textbf {\bibinfo {volume} {99}},\ \bibinfo
  {pages} {155001} (\bibinfo {year} {2007})}\BibitemShut {NoStop}%
\bibitem [{\citenamefont {Frenkel}\ and\ \citenamefont
  {Smit}(2000)}]{smit:MinimumImage}%
  \BibitemOpen
  \bibfield  {author} {\bibinfo {author} {\bibfnamefont {D.}~\bibnamefont
  {Frenkel}}\ and\ \bibinfo {author} {\bibfnamefont {B.}~\bibnamefont {Smit}},\
  }\href@noop {} {\emph {\bibinfo {title} {Understanding Molecular Simulation:
  From Algorithms to Applications}}}\ (\bibinfo  {publisher} {Academic Press,
  San Diego},\ \bibinfo {year} {2000})\BibitemShut {NoStop}%
\bibitem [{\citenamefont {Hockney}\ and\ \citenamefont
  {Eastwood}(1981)}]{eastwood:SimulationTheory}%
  \BibitemOpen
  \bibfield  {author} {\bibinfo {author} {\bibfnamefont {R.}~\bibnamefont
  {Hockney}}\ and\ \bibinfo {author} {\bibfnamefont {J.}~\bibnamefont
  {Eastwood}},\ }\href@noop {} {\emph {\bibinfo {title} {Computer Simulation
  using Particles}}}\ (\bibinfo  {publisher} {McGraw-Hill, New York},\ \bibinfo
  {year} {1981})\BibitemShut {NoStop}%
\bibitem [{\citenamefont {Berges}\ \emph {et~al.}(2004)\citenamefont {Berges},
  \citenamefont {Borsanyi},\ and\ \citenamefont
  {Wetterich}}]{wetterich:Prethermalization}%
  \BibitemOpen
  \bibfield  {author} {\bibinfo {author} {\bibfnamefont {J.}~\bibnamefont
  {Berges}}, \bibinfo {author} {\bibfnamefont {S.}~\bibnamefont {Borsanyi}}, \
  and\ \bibinfo {author} {\bibfnamefont {C.}~\bibnamefont {Wetterich}},\
  }\href@noop {} {\bibfield  {journal} {\bibinfo  {journal} {Phys. Rev. Lett.}\
  }\textbf {\bibinfo {volume} {93}},\ \bibinfo {pages} {142002} (\bibinfo
  {year} {2004})}\BibitemShut {NoStop}%
\bibitem [{\citenamefont {Gring}\ \emph {et~al.}(2012)\citenamefont {Gring},
  \citenamefont {Kuhnert}, \citenamefont {Langen}, \citenamefont {Kitagawa},
  \citenamefont {Rauer}, \citenamefont {Schreitl}, \citenamefont {Mazets},
  \citenamefont {Smith}, \citenamefont {Demler},\ and\ \citenamefont
  {Schmiedmayer}}]{schmiedmayer:Prethermalization}%
  \BibitemOpen
  \bibfield  {author} {\bibinfo {author} {\bibfnamefont {M.}~\bibnamefont
  {Gring}}, \bibinfo {author} {\bibfnamefont {M.}~\bibnamefont {Kuhnert}},
  \bibinfo {author} {\bibfnamefont {T.}~\bibnamefont {Langen}}, \bibinfo
  {author} {\bibfnamefont {T.}~\bibnamefont {Kitagawa}}, \bibinfo {author}
  {\bibfnamefont {B.}~\bibnamefont {Rauer}}, \bibinfo {author} {\bibfnamefont
  {M.}~\bibnamefont {Schreitl}}, \bibinfo {author} {\bibfnamefont
  {I.}~\bibnamefont {Mazets}}, \bibinfo {author} {\bibfnamefont
  {D.}~\bibnamefont {Smith}}, \bibinfo {author} {\bibfnamefont
  {E.}~\bibnamefont {Demler}}, \ and\ \bibinfo {author} {\bibfnamefont
  {J.}~\bibnamefont {Schmiedmayer}},\ }\href@noop {} {\bibfield  {journal}
  {\bibinfo  {journal} {Science}\ }\textbf {\bibinfo {volume} {337}},\ \bibinfo
  {pages} {1318} (\bibinfo {year} {2012})}\BibitemShut {NoStop}%
\bibitem [{\citenamefont {Spitzer}(1967)}]{spitzer:LSTheory}%
  \BibitemOpen
  \bibfield  {author} {\bibinfo {author} {\bibfnamefont {L.}~\bibnamefont
  {Spitzer}},\ }\href@noop {} {\emph {\bibinfo {title} {Physics of Fully
  Ionized Gases}}}\ (\bibinfo  {publisher} {Interscience, New York},\ \bibinfo
  {year} {1967})\BibitemShut {NoStop}%
\end{thebibliography}
\end{document}